\documentclass[12pt,aps,prd,preprint,showkeys]{revtex4}
\usepackage[utf8x]{inputenc}
\usepackage{amsmath}
\usepackage{amsfonts}
\usepackage{amssymb}
\usepackage{graphicx}

\begin{document}
\title{ Quark cores in extensions of the MIT Bag model}
\author{Salil Joshi}
 \affiliation{School of Physics, University of Hyderabad, Gachibowli, Hyderabad, India 500046}
 \author{Sovan Sau}
 \affiliation{School of Physics, University of Hyderabad, Gachibowli, Hyderabad, India 500046}
\author{Soma Sanyal}
\affiliation{School of Physics, University of Hyderabad, Gachibowli, Hyderabad, India 500046}

\begin{abstract}

Recent observations of massive pulsars having masses of the order of two solar mass pose a new challenge for compact objects such as hybrid stars and neutron stars. Extensions of the bag 
model and the Nambu-Jona-Lasino model have been used to model these stars to get higher mass stars. Quark matter has been predicted in the cores of these massive stars.  
In this work we show that an extension of the bag model, with a chemical potential dependent 
bag parameter can lead to an isentropic phase transition in the core of the neutron star. Our model shows that an EoS having all three quarks $u$, $d$ and $s$ would lead to massive stars with stable quark matter. We find that the mass of the stars not only depends on the bag constant but also on the mass of the strange quark. The mass - radius ratio which determines the redshift values on the surface, indicates that it is possible to obtain  stable self bound strange quark matter stars with reasonable values of the bag pressure which correspond to the recent observations of large mass stars.


\end{abstract}

\maketitle

\section{Introduction}

Among the various hypothetical compact objects discussed in the  literature, the quark star and the hybrid star hold a very important place, as they are objects capable of generating gravitational waves. The detection of gravitational waves by recent experiments raise the possibility of experimental determination of these special stars. Stable quark stars had been first predicted in ref.\cite{witten}; this was followed by various other studies on quark stars \cite{quarkstars}. The possibility of quark matter inside the cores of neutron stars has also been discussed \cite{Collins,Baym,alvarez}. Recent observations of relatively massive pulsars like PSR J1614-2230 and PSRJ0348+0432, have further contributed to the challenges presented by these stars. 

The Equation Of State (EoS) of these stars  present a challenge in the current scientific scenario. The 
most commonly used EoS is the bag model \cite{bagmodel,bagmodel1,bagmodel2,bagmodel3}. Apart from the standard bag model there have been discussions in the literature of several extensions of the bag model \cite{extensions}.
Recent studies of hybrid stars have shown that it is possible to have stars with masses of the order of
two solar masses and higher. These higher mass limits are usually set by observations of pulsar glitches. The masses of these stars  (PSR J1614-2230 and PSRJ0348+0432) \cite{star1, star2}  can be successfully modelled in extensions of the bag model\cite{ritammalik}. In these studies, the author had used a density dependent bag constant. Such density dependent bag parameter has also been used by other groups to study strange stars \cite{nprasad}. Various modified bag constants have been used to understand the quark hadron phase transition in baryon dense environments. It has been established that bulk strange matter can be stable \cite{farhi}. They can be formed by the trapping of strange matter due to phase transitions in the early universe or by the slow burning of the core of the neutron star which gradually changes it to a quark star \cite{harko}. Thus strange bulk matter can exist as cores in neutron stars, as quark stars or as hybrid stars. Apart from these there have been discussions of a color flavor locked phase in the core of the neutron stars and strange neutron stars \cite{cfl}. The Nambu Jona Lasino model has also been used to show that the core of massive neutron stars may contain a region of quark hybrid mass \cite{njl}. All these models have been motivated by various observational data on the mass and radius of these stars.    

There is also the question of the phase boundary between the confined and the deconfined phases. The nature of the quark hadron phase transition usually gives the kind of boundary between the two phases. In recent times, there have been quite a few models where a fixed entropy per baryon has been suggested at the phase boundary \cite{mariani}. The different stages of cooling of proto-neutron stars are modelled as isentropic phase transitions. This has been done using the Maxwell construction. Other than this, the isentropic phase transition is also considered for neutron stars at birth using detailed simulations \cite{masuda}. The isentropic quark-hadron phase transition had been considered for the extended bag model previously in the case of relativistic heavy ion collision as well as the early universe \cite{leonidov,patra}. It has been conjectured that an extended bag model with a temperature and baryon chemical potential dependent bag pressure can give an isentropic phase boundary between the two phases. Such modified bag models have been discussed in some detail in the literature . In this work we use the method 
introduced in ref. \cite{leonidov} by Leonidov et. al. to see whether such extension of the bag models can lead to stable strange bulk matter with mass and radius consistent with experimental observations. Our aim is to demonstrate that an extension of the bag model can explain the current observables related to large mass stars with a deconfined quark gluon plasma at its core.  

Here we have characterized the properties of these massive objects using the bag constant and the mass of the strange quark. Though we have studied other parameters too, these two properties are the ones that we focus on. Since the strange quark mass is already constrained from other experiments, this helps in obtaining very specific and limited ranges for our compact objects. The quark matter that we get in this model is energetically favourable over the $^{56}$ Fe crystal. The other requirement of any such model is the gravitational stability. The standard way of checking the stability is through the Tolman - Oppenheimer - Volkoff (TOV) equations. The partition function usually gives the energy density, pressure and number density of the quark gluon plasma.
These are then substituted in the TOV equations.  The TOV equations are routinely used to study the stability of strong gravitational objects like the neutron stars, the hybrid stars and the quark stars.  The TOV equations corresponding to the EoS of the modified bag model are then solved numerically to obtain the mass radius ratio corresponding to various different chemical potentials at different temperatures. We find that the EoS leads to structures that are quite stable for a large range of chemical potentials at low temperatures. 

We also study an experimental observable that is directly related to the compactness of the star. The mass and radius of the star allow us to calculate the surface red shift of the star. This can be measured through direct observation and is well documented for most stars. There are several massive stars which are considered as candidates for strange stars. We calculate the red shift from our model and find that it is within the range of several candidates for strange stars. We will discuss these candidates in detail in section IV of the paper.

In section II we present the EoS of the extended bag model used for modelling the quark cores of the strange matter in more detail. We have divided this section into three subsections. In the first subsection, we discuss the bag model EoS with two massless flavors of quarks, the $u$ and the $d$. In the second subsections, we discuss the EoS with three massless flavors of quarks, the $u$, $d$ and the $s$. In the third subsection we discuss the EoS where the strange quark is taken to be massive compared to the other two quarks. In section III, we discuss the TOV equations and their numerical solutions, In section IV, we present the results and discuss the possible candidates which fit our model. Finally we present our conclusions in section V. 

\section{The extended bag model}

We first briefly discuss the extended bag model that we use in our calculations. The original MIT bag model was the first phenomenological model which successfully modelled the phase transition of the quark gluon plasma (QGP) state to the hadronic phase. The phase transition was a first order phase transition which led to inhomogeneous baryon densities in the plasma \cite{witten}. The equation of state consisted of a bag constant $B$ which was actually the pressure difference between the two phases. Leonidov et. al \cite{leonidov} modified the bag model by changing the bag constant to a variable dependent on chemical potential $\mu$ and temperature $T$. Their model was consistent with the Gibbs equilibrium criteria for a phase transition. The model also conserved baryon number and entropy at the phase boundary. Following Leonidov's approach others also modified the bag constant by making it dependent on the chemical potential and temperature \cite{patra}. The discovery of the massive neutron stars indicates that the EoS leading to the quark cores in the neutron stars has to be reasonably stiff. We find that the isentropic phase transition in the core with a chemical potential and temperature dependent bag constant does provide this stiffness to the EoS.   
The entropy per baryon number is also maintained at the boundary. We find that it is possible to have self bound strange quark matter in the core within reasonable values of bag pressure. 
Though an isentropic phase transition is not a strong requirement for phase transitions in these conditions, the grand canonical partition function used by Leonidov \cite{leonidov} and Patra \cite{patra} do generate the stiffness that is required to have stars with a mass greater than  $1.5 M_0$ and radii of the order of $12$ kms.    

 We use their grand canonical partition function and  calculate the number density and the pressure of these stars. For massless quarks, both two flavors ($u$ and $d$) as well as three flavors ($u$, $d$ and $s$),  the energy density and pressure is given by,  
\begin{equation}
\begin{split}
    \epsilon=\frac{N_c N_f}{2}(\frac{7}{30}\pi^2 T^4+ \mu^2T^2 +\frac{\mu^4}{2\pi^2})+\frac{\pi^2}{15}N_g T^4 +  \\ 
    B(\mu,T)-\frac{\partial B(\mu,T)}{\partial \mu}T -\frac{\partial B(\mu,T)}{\partial T}\mu
 \end{split}   
\end{equation}
Here $N_c$ and $N_f$ are the number of color and flavor degrees of freedom. 
\begin{equation}
\begin{split}
    P=\frac{N_c N_f}{6}(\frac{7}{30}\pi^2 T^4+ \mu^2 T^2  + \frac{1}{2\pi}\mu^4)+ \\ \frac{\pi^2}{45}N_g V T^4 - B(\mu,T)
  \end{split}  
\end{equation}

In our model, we consider the EoS in the limit of high baryon density. As was shown by Leonidov \cite{leonidov}, in this limit, the thermodynamical quantities can be written as the sum of two parts. One part gives the zero temperature contribution and the second part gives the finite temperature contribution. When we consider the $s$ quark to be massive, these contributions change significantly. Hence in the next subsections, we will first consider the quarks to be massless and obtain the mass and radius relations for the three flavor case.  Then we will consider the strange quark to be massive and obtain the mass and radius relations for that case separately. For the two flavor case, the plasma is found to be meta stable.  

\subsection{Case 1: Two massless flavors ($u$ and $d$)}

We first look at the two flavor $u$ and $d$ case. Here we do not consider the leptons but maintain charge neutrality by assuming that the number of $d$ quarks is twice that of the $u$ quarks. As has been shown in ref.\cite{madsen}, the plasma stability
parameters are only weakly dependent on lepton contributions in the case of the two flavor plasma hence the lepton contribution to the charge neutrality can be ignored.  The general way to construct the  EoS is to describe the EoS in each of the two phases and then
perform the Maxwell construction to join the two phases along their common boundary.
We follow Leonidov's approach in modifying the bag constant.  This means that at the phase boundary the following equation must hold, 
\begin{equation}
\frac{S_q - \frac{\partial B(\mu, T)}{\partial T}}{n_q - \frac{\partial B(\mu, T)}{\partial \mu}} = \frac{S_H}{n_H}
\end{equation}
Here the suffix "q" denotes the thermodynamical variables in the quark phase, while the suffix "H" denotes the same variables in the hadronic phase. 
The modified bag's constant in our case is the following, 
\begin{equation}
B(\mu,T) \simeq B_0 +\mu^2 T^2-\frac{\mu^4T^2}{\theta_H^2}
 \end{equation}
 where $\theta_H =(\mu^2-m_H^2)^{1/2}$  and $m_H$ is hadron mass.
 
While the core consists of the quarks, the other side of the boundary is the hadronic phase. In this case, the hadronic phase consists of a non-interacting neutron - pion gas. The hadronic partition function, is based on a hard core neutron neutron repulsion. Due to this, it is necessary to divide any thermodynamical quantity for point like particles by a volume factor \cite{leonidov}. The volume factor is given by, $(1+ \frac{4}{3} \pi r_n^3)$, where $r_n$ is the radius of the neutrons. For the case of the two flavor massless quarks, a change in the volume factor does not give any significant change in the results. However,in the case of three flavor massless quarks, a change in the volume factor affects the values of the chemical potential of the quarks and the value of the corresponding bag constant for which the star is stable. The important quantity that can bring in a change in the volume factor is the radius of the neutron. We will discuss this later when we discuss the three flavor massless case. As mentioned in the introduction, we need to check the gravitational stability of the massive objects. For this we need to solve the TOV equations using the pressure obtained from our EoS. So the pressure, in our case turns out to be, 
\begin{equation}
  P=\frac{1}{3}(\epsilon- 4 B_0)-\frac{1}{3} \left(\frac{2\mu^4 T^2}{\theta_H^2}-\frac{18\mu^6 T^2}{\theta_H^4} \right).
 \end{equation}

Apart from the gravitational stability, to form a stable plasma state the energy per baryon number of the quark gluon plasma has to be calculated.  For the two flavor quark matter at low pressures, the energy per baryon number should be larger than that of the nuclear plasma (which is around $940$ MeV) and also a $^{56}Fe $ crystal which has the minimum binding energy per nucleon \cite{madsen, olinto}. Substituting all  these constraints, we find the chemical potential  is constrained between  $313.5$ MeV to $331$ MeV. The bag constant is also constrained within a narrow range between $149$ MeV to $154$ MeV. The temperature does not have much effect on these constraints as long as it is below $1$ MeV. The maximum size, can be obtained for the minimum bag constant and highest value of chemical potential. The baryon density turns out to be $0.32/ fm^3$. However, the plasma in this case is meta stable and one cannot give a definite mass and radius to this plasma,  hence we  introduce the strange quark and look at the plasma composed of three massless quark flavors.   

\subsection{Case 2:  Three massless flavors ($u$, $d$ and $s$) }
We now consider the plasma to have  three massless flavors of quarks. We treat the plasma as a degenerate Fermi gas with equal numbers of quarks of different flavors. This and the fact that the plasma is beta equilibrated gives the necessary charge neutrality to the plasma. The only change in the expressions for 
the bag constant and the pressure come from the fact that the number of flavors have been increased to three. 
As of now the assumption of $\mu_u = \mu_d = \mu_s $ holds as all the flavors are considered to be massless. The bag parameter becomes, 
\begin{equation}
     B(\mu,T) \simeq B_0 +\frac{3}{2} \left(\frac{1}{9}\mu^2T^2-\frac{1}{81}\frac{\mu^4T^2}{\theta_H^2}\right)
\end{equation}
and the corresponding pressure equation is given by, 
\begin{equation}
    P=\frac{1}{3}(\epsilon- 4B_0)-\frac{1}{2}\left(\frac{2\mu^4T^2}{\theta_H^2}+\frac{18\mu^6T^2}{\theta_H^4}\right)
 \end{equation}
Here $B_0$ is the bag parameter at $T=0$. For this case too, we have to check the stability of the plasma. Since the number of flavors has increased, the energy per baryon will also change accordingly. The preferred state of a stable plasma comprising of all the quark flavors should have energy per baryon less than that the nuclear energy of $940$ MeV so that the plasma does not hadronize. Similarly, the energy per baryon should be less than the binding energy of the $^{56}Fe $ crystal. The lower energy per baryon will make the quark gluon plasma the preferred state for the bulk matter \cite{madsen,olinto}. As mentioned before, the stability of the plasma puts a constraint on the values of the chemical potential as well as the bag constant. We see that the stability of the plasma increases with smaller bag constant values.

Again similar to the previous case, since the hadronic phase is also present, there are some parameters in the hadronic phase which also contribute to the stability of the plasma. We find that in the three flavor case, the radius of the neutron which contributes to the volume correction in the neutron - pion gas does change the stability values of the core pressure. This is because we have equated the pressure of the hadronic phase and the quark phase to obtain the bag constant.      
We have plotted the pressure in the core for different values of bag constant for different values of the neutron radius for which the quark plasma is stable in fig 1. As can be seen from the figure, the increasing bag constant decreases the pressure in the core.

\begin{figure}
\includegraphics[width= 0.5 \textwidth]{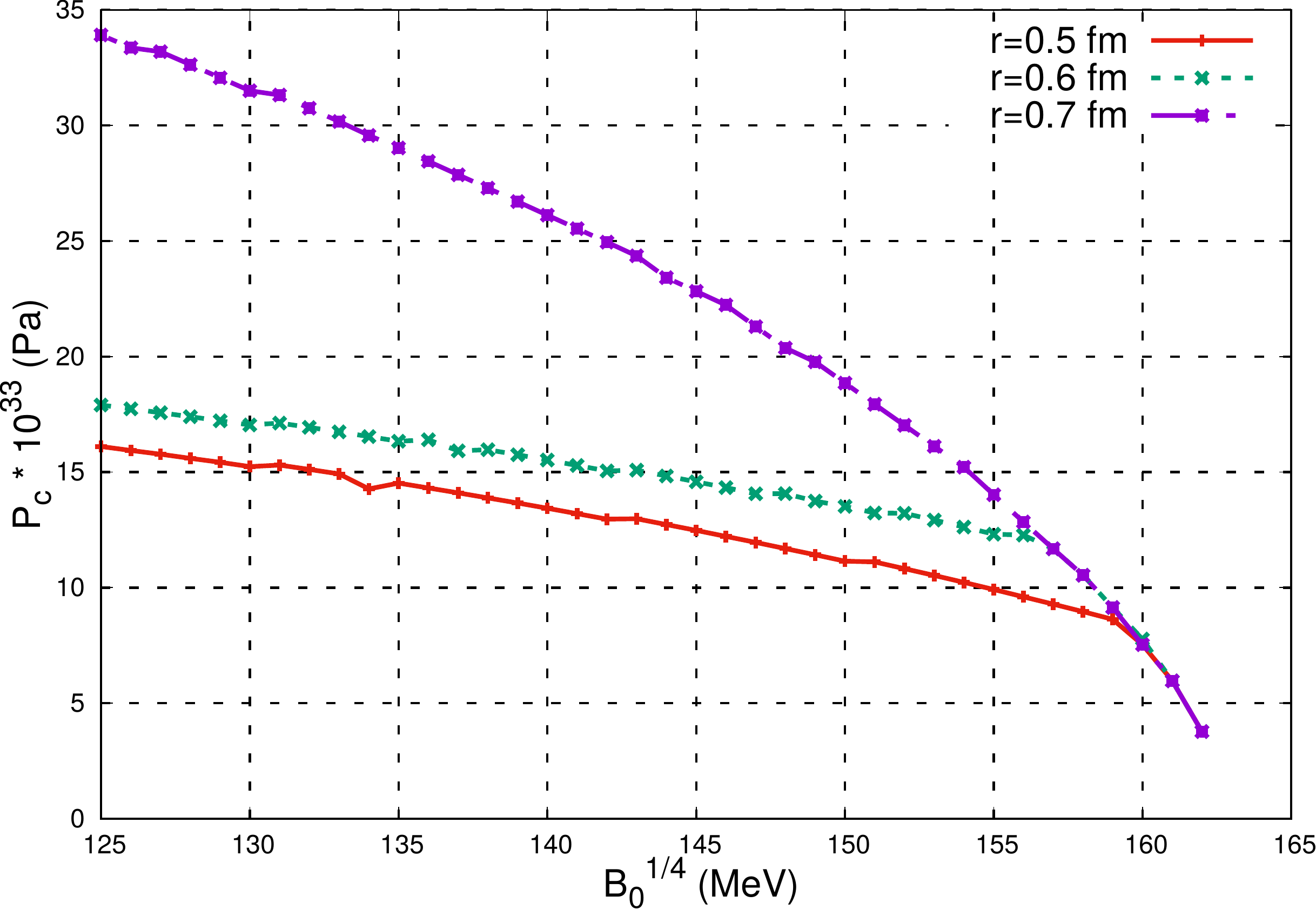}
\caption{Core pressure for stable masses at different bag constants for different values of the neutron radius. }
\end{figure}

We thus find that the change in the volume corrections to the non-interacting neutron-pion gas affect the limiting values of the pressure and the density at the core. In all the cases, we have considered we have kept the hadronic phase to be the same as we are interested in the quark phase. We find that as long as the hadronic part consists of a neutron - pion gas, the only parameter that affects the stability of the plasma is the radius of the neutron through the volume corrections. Though this effect also occurs when the strange quark mass is taken into account, the dominance of the strange quark mass leads to very small variations in the core pressure due to the change in the volume corrections. 

\subsection{Case 3: Massless $u$,$d$ and massive $s$ quarks}

If we consider the $s$ quark to be massive then we cannot have $\mu_u = \mu_d = \mu_s $. So it is important to consider the first order correction terms to the temperature and the entropy. As in the previous case, strange matter is modelled as a Fermi gas of up, down, and strange quarks. This time the charge neutrality of the system is maintained by the electrons. 
Weak interactions ($\beta$ equilibrium) will then lead to $\mu_s = \mu_d = \mu $ and $\mu_u + \mu_e = \mu $ \cite{alcock}. The number densities are related by, $\frac{2}{3} n_u - \frac{1}{3}n_u - \frac{1}{3}n_s - n_e = 0$.  
This means that effectively we are dealing with only two chemical potentials $\mu_u $ and $ \mu_s$.

To deal with high baryon densities and low temperatures, finite temperature corrections are added to the zero temperature terms. This means that, 
\begin{equation}
P_s \simeq P_s^0 + P_s^1 T^2 + P_s^2 T^4 
\end{equation}
\begin{equation}
n_s \simeq n_s^0 + n_s^1 T^2 + n_s^2 T^4 
\end{equation}
\begin{equation}
S_s \simeq   S_s^1 T + S_s^2 T^3 
\end{equation}
The zero temperature terms are given by,
\begin{equation}
P_s^0 = \frac{1}{6 \pi^2}(\mu_s \theta_s (\theta_s^2 - \frac{3 m_s^2}{2}) + \frac{3 m_s^4}{2} ln (\frac{\mu_s+\theta_s}{m_s}))
\end{equation}
\begin{equation}
n_s^0 = \frac{2 \theta^2}{9 \pi^2}
\end{equation}
Here $m_s$ is the mass of the $s$ quark and $\theta_s = \sqrt{(\mu_s^2 - m_s^2)}$. 
 At high baryon densities and low temperatures $\frac{\partial B}{\partial T}$ dominates over $\frac{\partial B}{\partial \mu}$, based on that at the phase boundary one can obtain the bag constant as, 
\begin{equation}
      B(\mu,T)\simeq B_0 +(\mu_u^2+\mu_d^2 - \frac{2}{3}\mu_s\theta_s)\frac{T^2}{2} - \frac{\mu_s T^2}{ 2 \theta_H^2}(\mu_u^3+\mu_d^3+\frac{2}{3}\theta_s^3)
\end{equation}

Once the bag constant is obtained the pressure can be expressed as,  
\begin{equation}
\begin{split}
P =\frac{1}{3}(\epsilon- 4B_0 \frac{4 \mu_e^4}{3 \pi^2} + \frac{m_s^4}{\pi^2}ln(\frac{\mu_s+\theta_s}{m_s})- (\mu_u^3+\mu_d^3 -\frac{1}{3}\theta_s^3)(\frac{\mu_s T^2}{\theta_H^2}) + \frac{\mu_s \theta_s}{2 \pi^2} (\theta_s^2 - \frac{3 m_s^2}{2}) \\ - \frac{ \mu_s^3 \theta_s T^2}{\theta_H^2} - \frac{\mu_s}{2 \pi^2} (\mu_s^2 \theta_s + \frac{m_s^2}{2 \theta_s (m_s^2 - \mu_s^2)}) + \frac{\mu_s T^2}{2} (\frac{18 \mu_s^2}{\theta_H^4} (\mu_u^3 + \mu_s^3 + \frac{2 \theta_s^2}{\theta_H^4})))
\end{split}
\end{equation}

As we can see apart from the chemical potentials of the $u$ and $s$ quark, the main parameters are the $s$ quark mass, the temperature and the bag constant. The $s$ quark mass is already constrained by other experiments. So though we do get large stable mass stars with a smaller value  of the $s$ quark mass, we do not vary the $s$ quark mass below $90$ MeV . Generally this is the lower limit of the $s$ quark mass from various other sources \cite{olive}. The temperature is also below $1$ MeV. The constraint on the temperature comes as the entropy per  baryon number has to be continuous even in the bulk. Increasing the temperature violates this continuity and hence we are constrained to temperatures below $1$ MeV.   
The electron fraction in the core increases in this temperature range reaching a maximum at $1$ MeV, consequently the strangeness fraction goes down. So we find that lower the temperature, higher is the strangeness fraction in the core. However, the strangeness fraction saturates to a value of $0.33$ and becomes independent of temperature at lower values.   

In the previous section, we had seen that some parameters in the hadronic phase also affect the stability of the quark cores. We had found that changes in the volume corrections to the neutron - pion gas cause variation to the core pressure. In this case, the chemical potential of the quarks $u$ and $s$ along with the mass of the strange quark dominate the core pressure. In fact as the chemical potential of the $u$ quark is increased, the electron fraction and the strangeness fraction required for a stable core goes down while the baryon number density  increases to maintain the stability. The rate of increase(decrease) of the electron fraction, the strangeness fraction and the baryon number density has been plotted against the chemical potential of the $u$ quark in fig 2. Generally a larger value of the chemical potential of the $u$ quark ($\mu_{u}$) results in more stable quark cores provided the strange quark mass is on the lower side. This is seen by calculating the energy density of the quark core, which has to be more stable than the $^{56} Fe$ nucleus. For all the values in fig 2, we get  stable quark matter. The stability is checked by calculating the energy per baryon in the plasma. 
\begin{figure}
\includegraphics[width= 0.5 \textwidth]{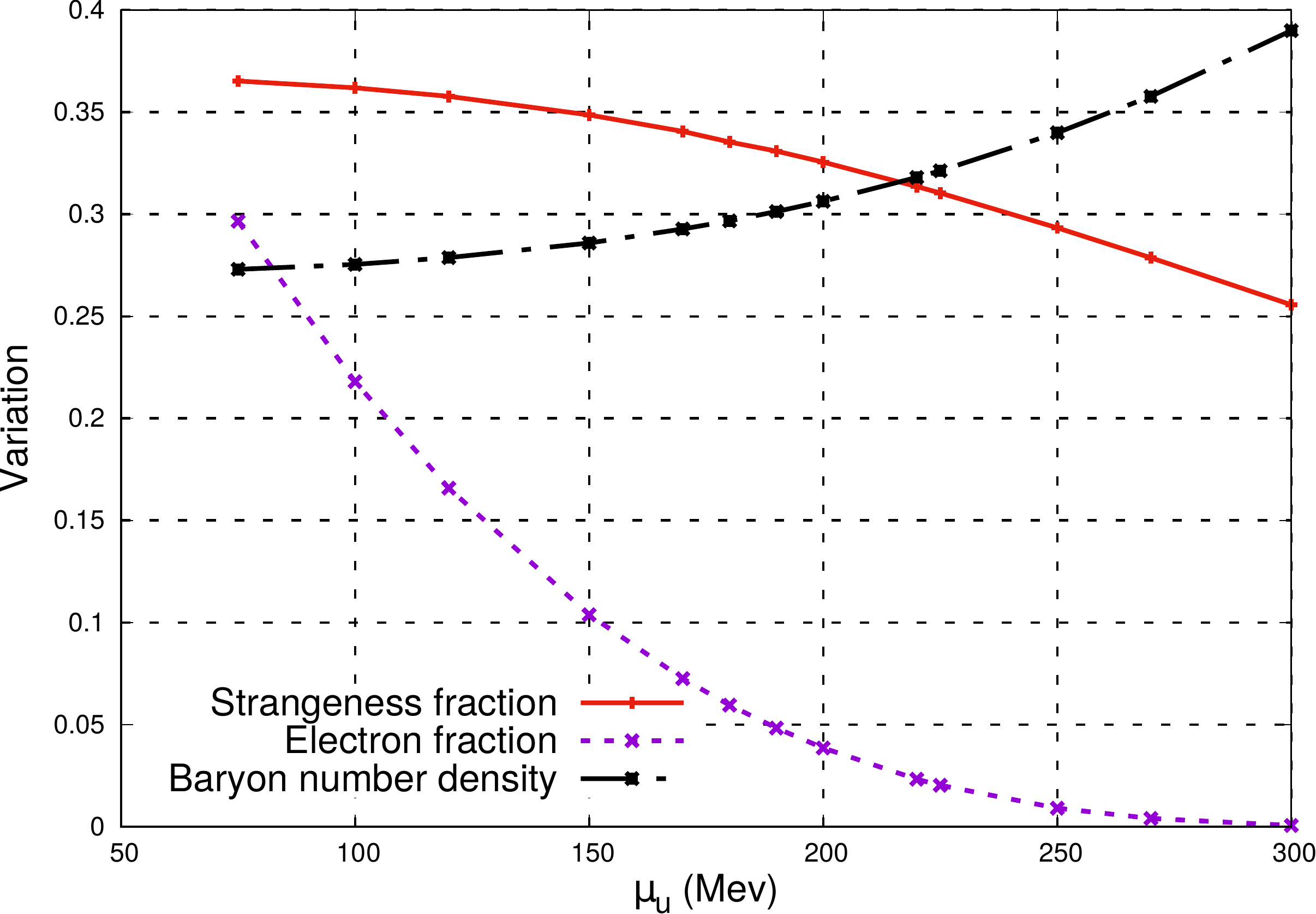}
\caption{The rate of change of the electron fraction, the strangeness fraction and the baryon number density with respect to the chemical potential of the $u$ quark. }
\end{figure}
Thus we have found that it is possible to have energetically stable quark matter using three flavors of the quark gluon plasma. As these are massive objects, we know that gravitation plays an important role in establishing the stability of these stars. The macroscopic quantities of mass and radius are thus calculated using the Tolman - Oppenheimer - Volkoff  equations \cite{TOV}. In the next sections, we briefly describe  these equations and their solutions.

\section{The Tolman - Oppenheimer - Volkoff  equations }
We consider the TOV equations for a spherical and isotropic metric. We assume that the core of the quark star has a uniform density in all the directions. The TOV equations are then given by,
\begin{equation}
\frac{dP}{dr} = \frac{G}{r}\frac{(\epsilon(r)+P(r))(M(r)c^2+4\pi r^3P(r)}{(rc^2-2GM(r))}
\end{equation}
and
\begin{equation}
\frac{dM}{dr} = \frac{4\pi r^2 \epsilon(r)}{c^2}
\end{equation}
where the mass density $\rho(r)=\frac{\epsilon(r)}{c^2}$ and $c$ (cm/sec) is the speed of light and $\epsilon(r)$ is the energy density of the plasma in MeV. The TOV
equations are solved numerically with appropriate boundary conditions to obtain the mass distribution of the stars. The
solution also gives us the gravitationally allowed values of the mass and radius of the stars.
We solve the TOV equations numerically for the high baryon density regime. We find that stable stars are possible for various values of the bag constant. Detailed results and graphs are shown in the next section.

Apart from the mass radius ratio, we would also like to look at a direct observable that is affected by the compactness of the star. One such observable is the surface redshift values of the stars. It is a parameter that corresponds to the redshift experienced by a radially propagating photon travelling from the star's surface to infinity. It can be directly calculated from the mass - radius ratio of a star.   
\begin{equation}
z_s = \left( 1 - \frac{2 G M}{R c^2} \right)^{-1/2} - 1
\end{equation}
For all the cases, where we have a stable configuration from plasma stability as well as gravitational stability, we have calculated the red shift values of the stars. In the next section, we will show that the surface redshift values obtained for the strange stars in our case are consistent with experimentally observed red shift values for different strange star candidates.   

\section{Results}

\subsection{Mass-radius ratios with three massless flavors ($u$,$d$,$s$)}
Since the two flavor plasma only gives us a metastable plasma we do not solve the TOV equations for the two flavor plasma. 
In the case of the three massless flavors of quarks we vary the various parameters and do a 
systematic search to put constraints on the various parameters of the model. 

\begin{figure}
\includegraphics[width = 0.5 \textwidth]{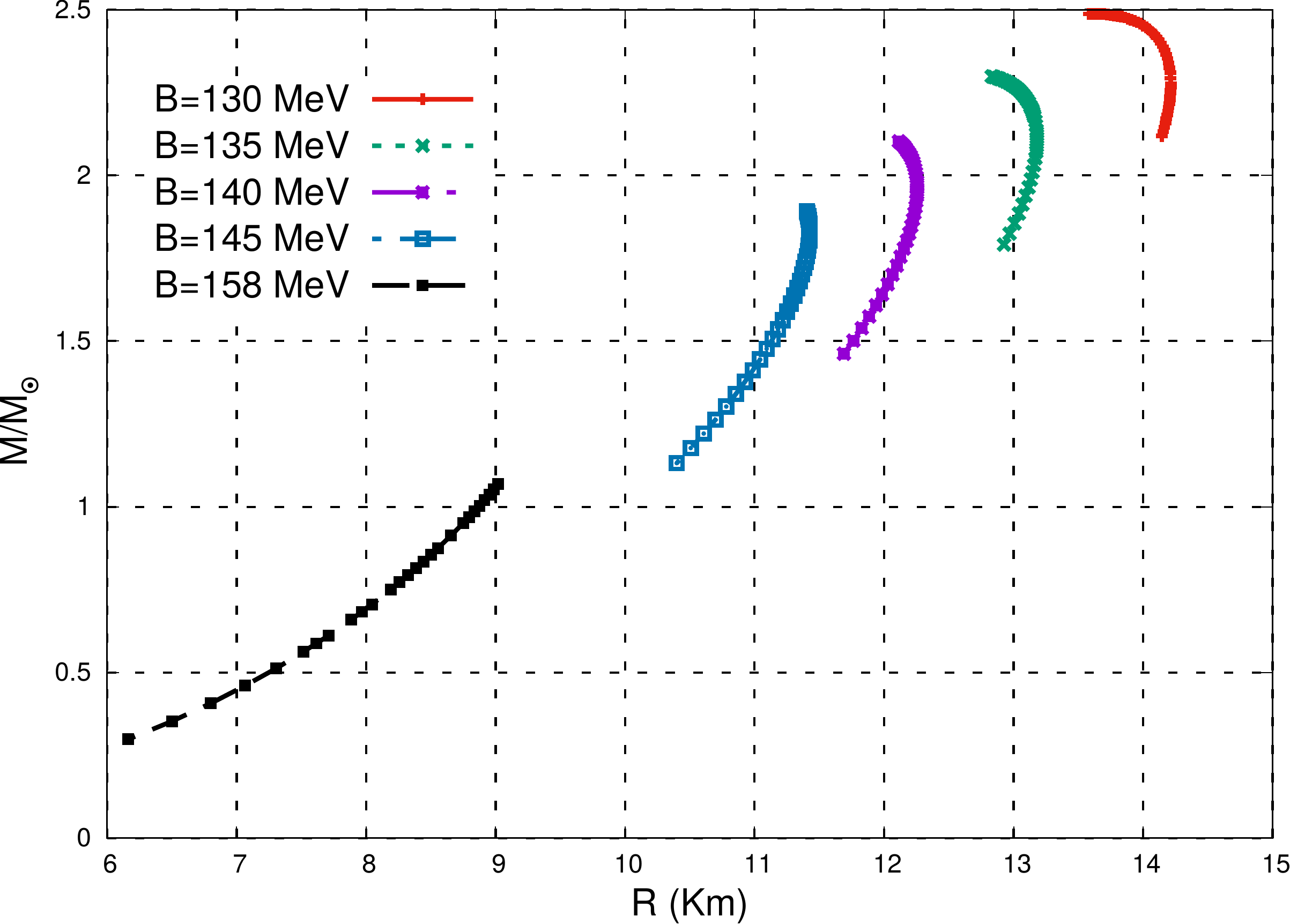}
\label{massless}
\caption{$M$ (in solar masses) Vs $R$ at different bag constants for massless quarks. }
\end{figure}
For massless quarks, we do get large mass values close to $2.2 M_0$ with a radius of $14$ km. This occurs for lower bag constants. The strange quark is considered massless here and therefore we get the mass values greater than two solar masses. 
We then obtain the surface red shift values for these large mass stars and the resultant plot is shown in fig 4.  
\begin{figure}
\includegraphics[width = 0.5 \textwidth]{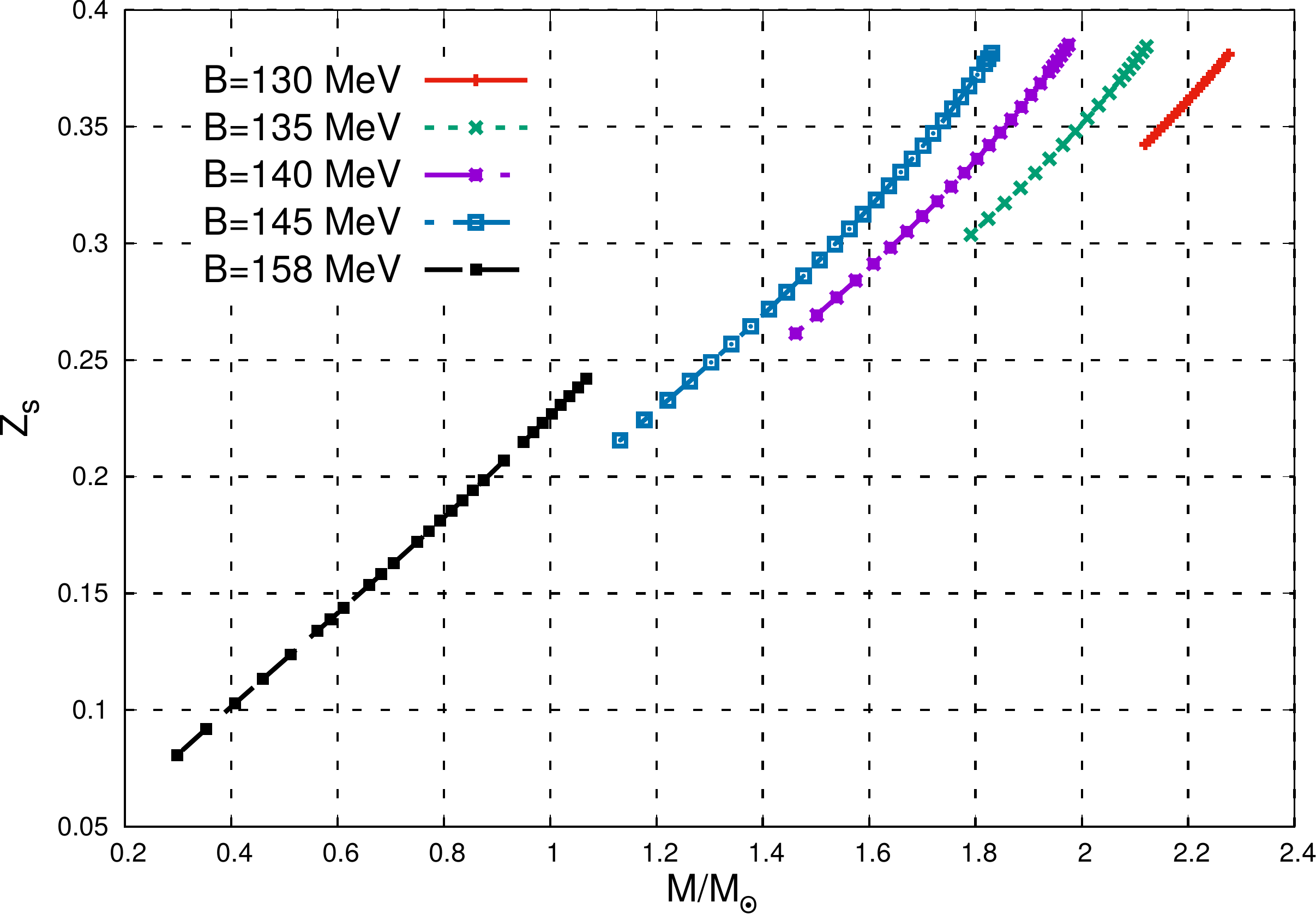}
\label{masslessredshift}
\caption{Red shift values for cores with massless quarks. }
\end{figure}
As seen from fig 4., we obtain redshift values of $0.4$ for massive stars of $2 M_0$ or higher. It has been predicted that PSR J0348+0432 has a radius of about $12$ kms and mass of $2.1 M_0$ \cite{star1} with a red shift value of $0.4$. So our model with massless quarks is able to reproduce the mass - ratio required to have stable quark stars as seen from the surface red shift data. To obtain the large mass stars, our bag constant should be around $130- 140$ MeV.

\subsection{Mass-radius ratios for the massive strange quark}

We now present the results of  the TOV equations with the pressure obtained from the EoS corresponding to the massive strange quarks. The mass of the quark puts a constraint on the quark chemical potentials. We plot the mass radius ratio for stable stars for two different masses of the strange quark. Since we know that the larger mass stars would occur in the range of bag constant values between $130 - 140$ MeV, we give the results for two values within this range. 

\begin{figure}
\includegraphics[width = 0.5 \textwidth]{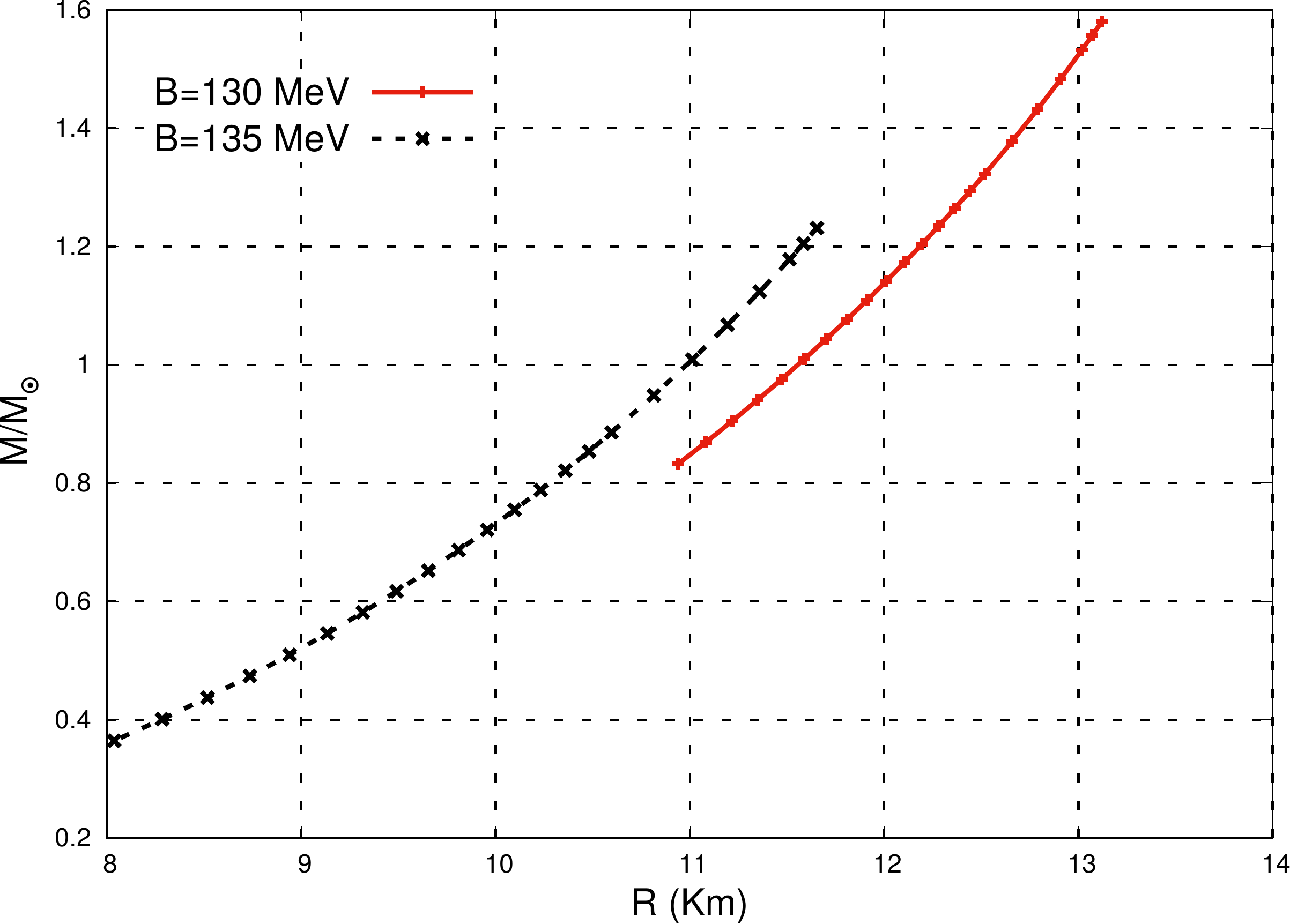}
\label{smlow}
\caption{ $M$ (in solar masses) vs $R$ at different bag constants for  $m_s= 100$ MeV, $\mu_u  = 100$ MeV for a u-d-s massive plasma }
\end{figure}

\begin{figure}
\includegraphics[width = 0.5 \textwidth]{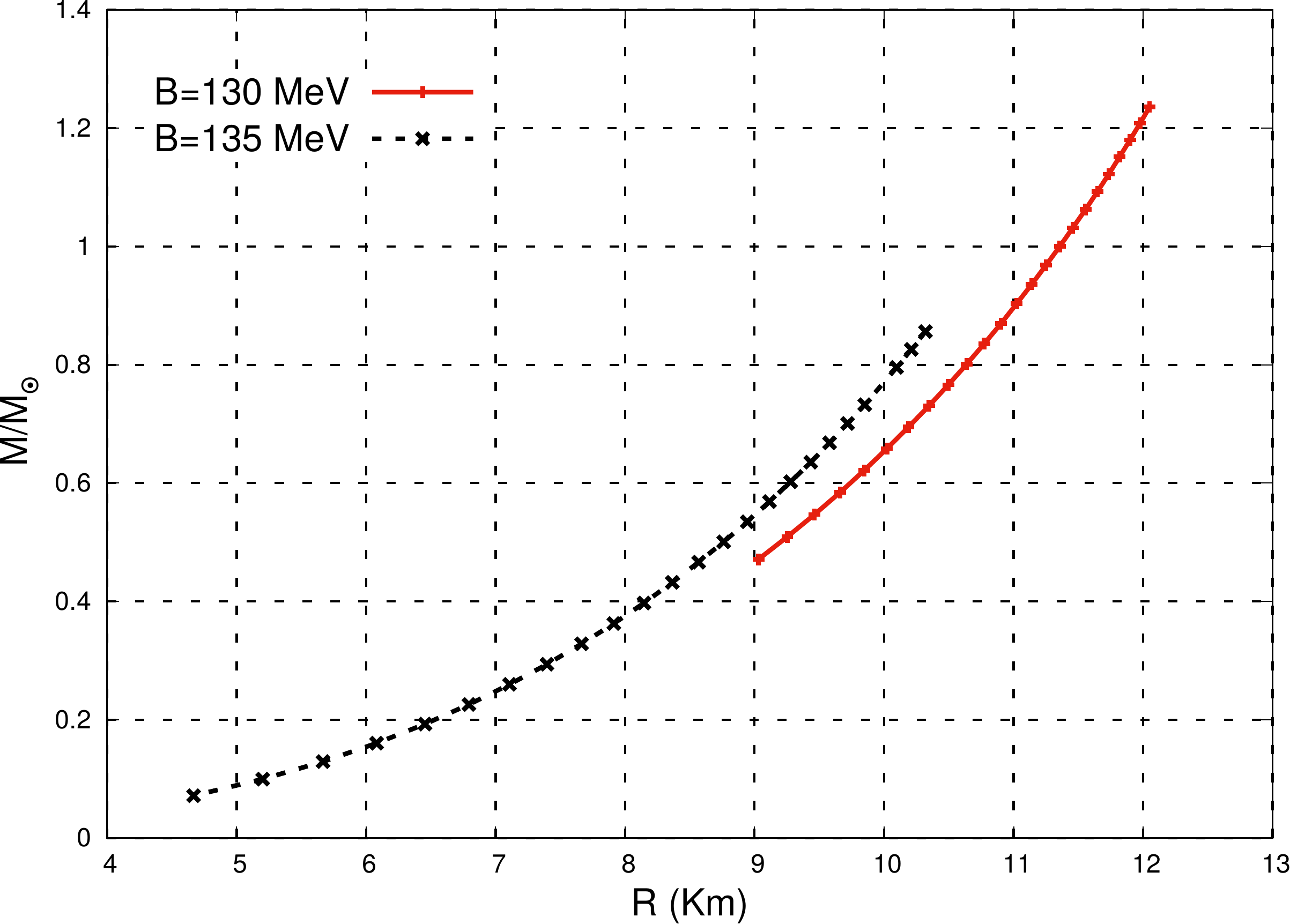}
\label{smhigh}
\caption{$M$ (in solar masses) vs $R$ at different bag constants for  $m_s= 150$ MeV, $\mu_u  = 100$ MeV for a u-d-s massive plasma }
\end{figure} 

As we see from fig 5  and fig 6, we get larger mass with smaller values of strange quark mass. So if the star has massive strange quarks, it's size will be constrained by the strange quark mass. We have plotted the redshift values for different masses of the stars.  

\begin{figure}
\includegraphics[width = 0.5 \textwidth]{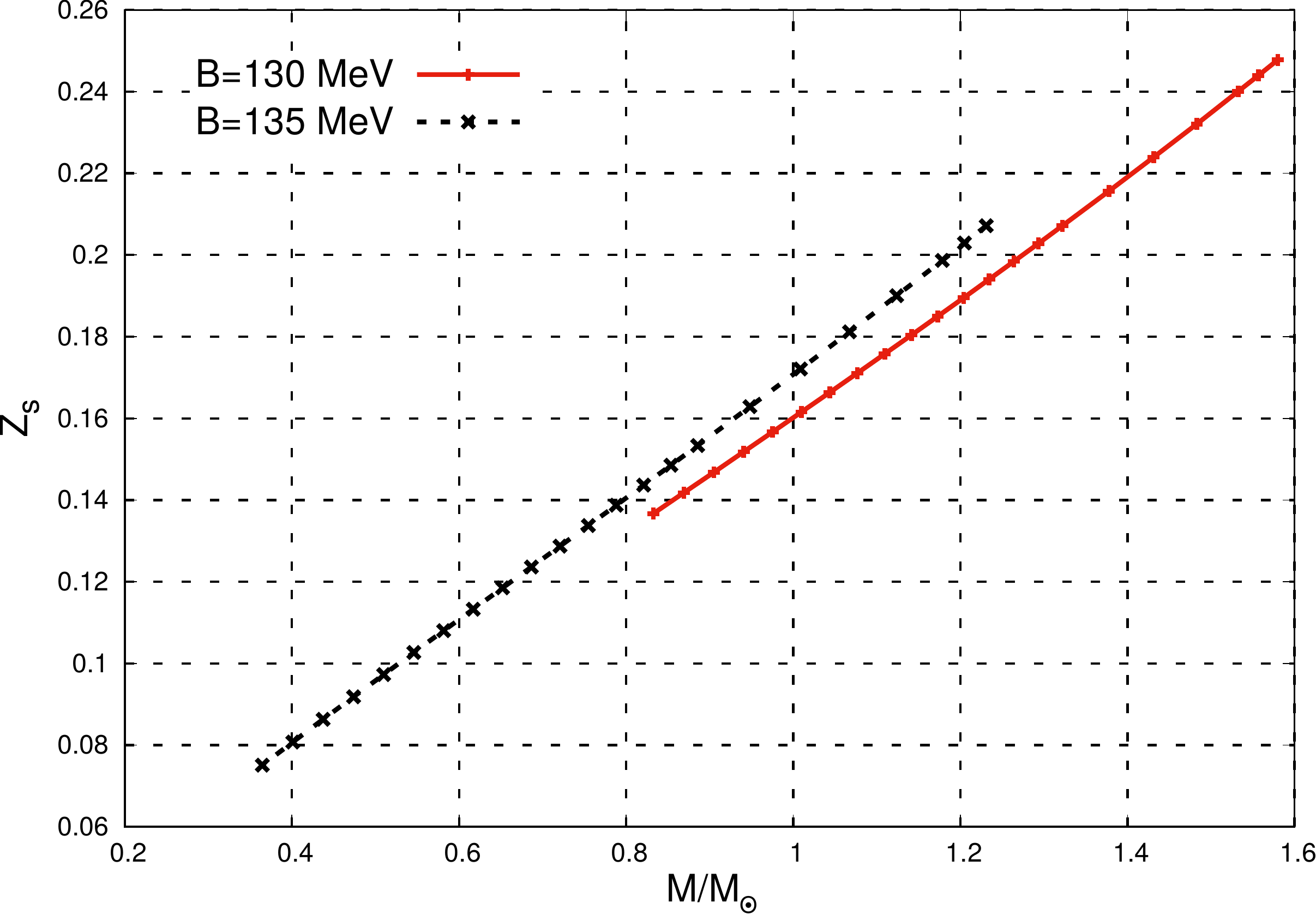}
\caption{Redshift values with the plasma having a massive strange quark of mass $m_s= 100$ MeV }
\end{figure}

\begin{figure}
\includegraphics[width = 0.5 \textwidth]{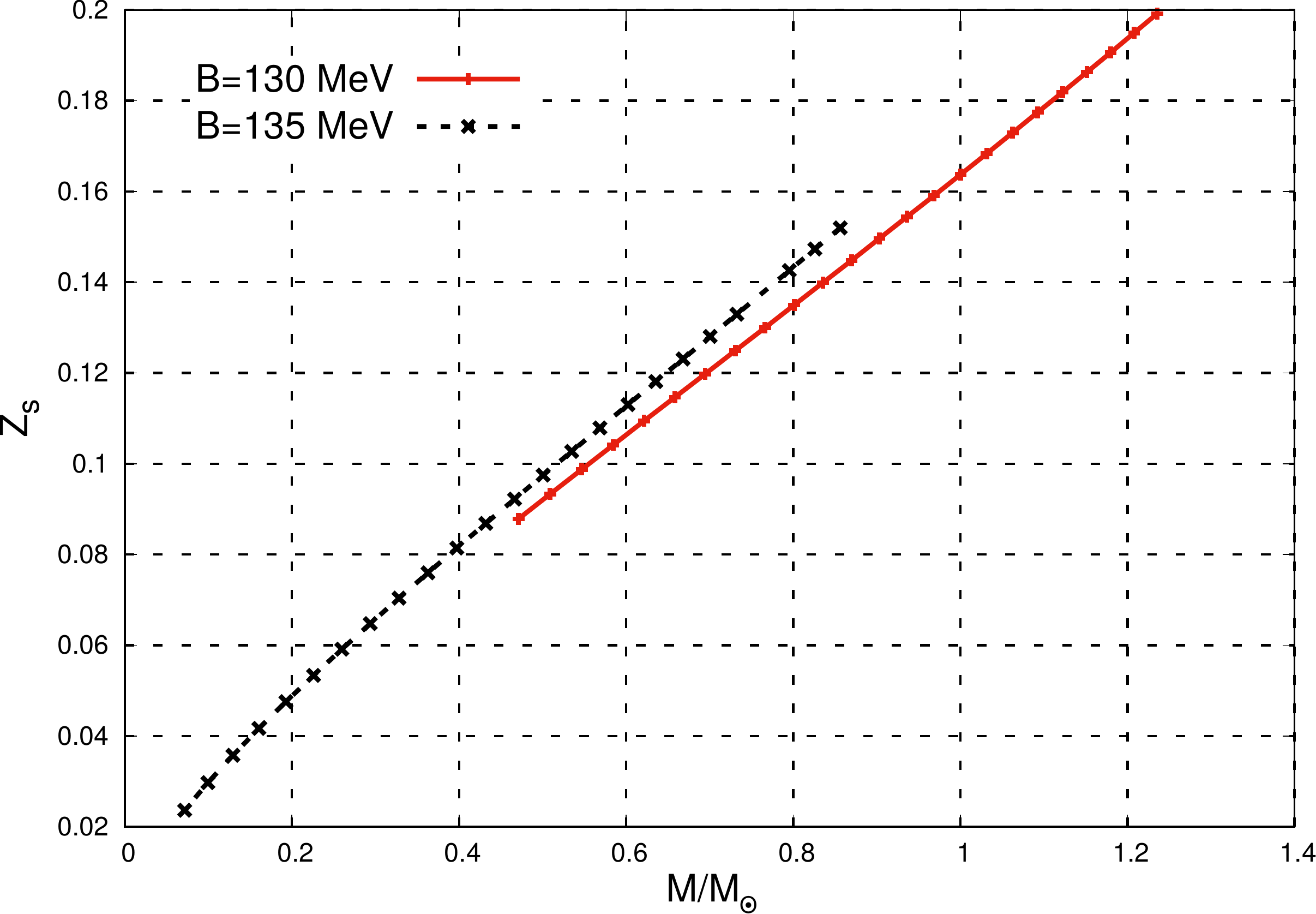}
\caption{ Redshift values with the plasma having a massive strange quark of mass $m_s= 150$ MeV}
\end{figure} 

Here since we have constrained the strange quark value to a lower limit of $90$ MeV, we do not get massive stars with masses greater than $2 M_0$. We can get such values if we lower the strange quark values to around $75$ MeV. We find that stars with masses of the order of $1.5 M_0$  would typically have surface red shift values around 0.22. This has also been seen in other models \cite{star1, star2}. Thus the observational constraints given by red - shift values are all satisfied with different values of strange quark mass in our model. We now discuss some observational data in support of our model. 

There are many observational instances of stars with these kind of parameters. The XTE J1739-285 with a mass of $1.51$ solar mass and 10.9 km radius \cite{star3}, is often referred to as a quark star. It is  rapidly spinning and has a  red shift ranging from $1.8$ to values of $2.4$ depending on its radius. Our model for the quark star with a massive strange quark of $100$ MeV and bag constants between $130 - 135$ MeV gives red shifts in this range. The EXO 0748-676 with a mass of $2.1$ solar masses and $13.8$ km in radius \cite{star4} has a red shift value of $0.35$. The EXO 0748-676 can be a quark star with a bag constant of $135$ MeV as per this model. For stars like PSR J1614-2230, the gravitational redshift values are in the range $0.41 - 0.5$, such high redshift values occur for the quark plasma with massless quarks. As we have mentioned before the mass - radius ratio and the red shift values depend crucially on the masses of the strange quark. 
 
\section{Conclusions}

In conclusion, we have looked at bulk strange matter using a bag constant which is dependent on chemical potential and temperature. The bag constant in this model had originally been derived by Leonidov et. al. for two massless flavors of quarks. In this work we have extended their model to include the strange quark. Though an isentropic phase transition is not essential in the core of the star, we show that such a phase transition can lead to stable quark matter inside the stars.  
We have done a systematic study of the parameters of the model. We have shown that the presence of the massless strange quark increases the stability of the plasma and the mass limits of these stars. The stability of the plasma limits our bag constant to the lowest value of $130$ MeV. As is already known, the stability of the quark matter increases with decrease of the bag constant. However, in this work, we have also shown that apart from the bag constant, it is the mass of the strange quark that determines the radius and mass of the massive object. Since the mass-radius ratio determines the surface redshift of the star, we have also obtained the surface red shift value for both massless and massive strange quark stars. We have shown that our model can reproduce the mass- radius ratio and the surface red shift of   
stars like PSR J1614-2230 and PSR J0348-0432 for bag constant values in the range $130 - 140$ MeV. The mass of the strange quark is important and we get larger mass stars for lower values of strange quark mass. Though we have maintained the lower value of strange quark mass at $90$ MeV due to constraints from other sources, there are previous studies where the strange quark mass is considered from $80$ MeV \cite{nanapan} onwards. Lowering the strange quark mass will result in higher mass stars. Thus we have established that it is quite possible to have a stable quark gluon plasma bulk phase in the core of a massive star with no other exotic phases.

Apart from PSR J1614-2230 and PSR J0348-0432  already mentioned before, both mass-radius ratios and the  red shifts of XTE J1739-285 and EXO 0748-676 can be explained by our model. Thus we do find a large number of candidates with larger masses and surface red shifts which fit our model. We have not considered the rotational effects of this model in this work, we have studied the stability of the quark mass only in the bulk. We have established that it is possible to model masses of the order of  two solar masses using extensions of the three flavour bag model with a $s$ quark whose mass essentially determines the mass-radius ratio of the star. Our future plan is to study the rotational effects of these stars.        

\begin{center}
 Acknowledgments
 
 S.J would like to acknowledge partial financial support (through long term project scheme) from the UGC Networking Resource Center, School of Physics, University of Hyderabad.  
\end{center}

\end{document}